\documentclass[twocolumn,showpacs,preprintnumbers,amsmath,amssymb,prb]{revtex4}
\usepackage{graphicx}
\usepackage{dcolumn}
\usepackage{bm}
\begin{document}
\title{Temperature dependence of the collective mode and its influence on
the band splitting in bilayer cuprates}
\author{S. Varlamov and G. Seibold}
\affiliation{Institut f\"ur Physik, BTU Cottbus, PBox 101344,
         03013 Cottbus, Germany}
\date{\today}

\begin{abstract}
The recently observed bilayer splitting in high-$T_{\rm c}$
cuprates is analyzed within a model where the charge carriers are
coupled to a phenomenological bosonic spectrum which interpolates
between the marginal Fermi liquid structure and collective mode
type behavior as a function of temperature. We argue that the
origin of the collective mode is associated with dynamic
incommensurate charge density waves. Moreover it is shown that the
resulting temperature dependence of the self-energy $\Sigma$ is in
good agreement with that extracted from angle-resolved
photoemission data.
\end{abstract}
\maketitle

\section{Introduction}

One of the long standing unsolved problems in angle-resolved
photoemission spectroscopy (ARPES) of high-$T_{\rm c}$ cuprates
was the detection of band doubling in bilayer materials. In these
compounds one naturally expects a splitting of the quasiparticle
(QP) dispersion in an antibonding (AB) and bonding band (BB) due
to finite hopping of charge carriers between the two CuO$_2$
planes. Only recently this feature has been measured in
Bi$_2$Sr$_2$CaCu$_2$O$_{8+\delta}$ (Bi2212)
\cite{FENG,CHUANG,DESSAU} above and below $T_{\rm c}$. The
splitting turns out to be essentially zero along the diagonals
$(0,0)\to(\pi,\pi)$ and (in the normal state) aquires a value
between $88meV$ \cite{FENG} and $110meV$ \cite{CHUANG} at the
$(\pi,0)$ point of the Brillouin zone (BZ). In addition it seems
that the magnitude of the energy splitting is constant over a
large range of doping. \cite{CHUANG2} Most interestingly it turns
out that for the same doping the splitting is only $22meV$ below
$T_{\rm c}$  \cite{FENG} contrary to the expectation that the
hopping between planes $t_{\perp}$ should be essentially
temperature independent. 

In this paper we show that the apparent
reduction of the bilayer splitting below $T_{\rm c}$ can be
explained by the assumption that the charge carriers are coupled
to bosons which frequency distribution evolves from a featureless
structure well above $T_{\rm c}$ (i.e. approaching that of a
marginal Fermi liquid \cite{VARMA}) to a sharp, mode-like
excitation at small temperatures. As a consequence, low energy QP
excitations are essentially confined to energies up to the mode
frequency $\omega_0$ which thus constitutes an upper limit for the
splitting between AB and BB QP peaks (see Fig. 4). Moreover, the
damping of excitations for energies larger than $\omega_0$ leads
to a broad hump which turns out to be mainly determined by the
bonding band (cf. Fig. 3c). This scenario is thus in agreement
with recent ARPES experiments \cite{DESSAU} where this peak-dip
hump structure at $(\pi,0)$ has been resolved below $T_{\rm c}$ in
addition to the bilayer splitting.

Numerous experiments now support the existence of a collective
mode in the high-$T_{\rm c}$ materials most prominently
exemplified by the appearance of a kink in the electron dispersion
along $(0,0)\to(\pi,\pi)$ as seen in ARPES \cite{VALLA,LANZARA}
(cf. also Fig. 2). Concerning the origin of the mode the various
proposals include a magnetic resonance (see e.g. Ref. \onlinecite{CAMP}), 
the coupling to phonons \cite{LANZARA} or incommensurate
charge-density waves (ICDW) \cite{GOETZ} as the source of the
associated scattering. It has been argued \cite{CAMP} that the
difference between peak and dip position of the photoemission line
shape at $(\pi,0)$ decreases as a function of doping as it is
expected for the magnetic resonance frequency. An analagous
analysis \cite{LANZARA} came to the conclusion that the frequency
is essentially constant (if not slightly decreasing) upon doping
thus resembling the behavior of a phonon mode. We have recently
fitted \cite{GOE2} ARPES data taking into account also the
feedback of the doping dependent superconducting (SC) gap. The
extracted mode frequency is a strongly decreasing function of hole
concentration reaching a value of $\omega \sim 40meV ... 50meV$
around optimal doping. This behavior is thus consistent with a
quantum critical point (QCP) located near optimal doping where the
associated order parameter corresponds to ICDW formation.
\cite{CAST} In fact, the formation of stripe order is now well
established in a large class of cuprate compounds (for an overview
see e.g. Ref. \onlinecite{PROCEED}). 
Most interestingly the coexistence of
charge modulations and superconductivity has recently also been
reported for the Bi2212 compound using scanning tunneling
microscopy. \cite{HOFFMAN,HOWALD} These experiments strongly
support our view that ICDW scattering can also be responsible for
the mode-like features as seen in ARPES experiments.
In the following we substantiate this scenario by a detailed analysis
of the temperature dependent collective mode
which allows us to fit the experimentally observed bilayer splitting in
the superconducting and normal state respectively by simply tuning
the broadening of our phenomoneological bosonic spectrum.

The paper is organized as follows. In Sec. II we
analyze available experimental data in order to elucidate the
frequency and temperature dependence of the collective mode in
Bi2212 materials. Based on these results   
the temperature dependent bilayer splitting is investigated within the
ICDW mode scenario in Sec. III.
We finally conclude our discussion in Sec. IV.

\section{Temperature dependence of the collective mode}

The coupling of charge carriers to a collective mode is reflected
in a drop of the imaginary part of the electronic self-energy Im
$\Sigma$ below some characteristic mode frequency $\omega_0$. In
Refs. \onlinecite{KAMIN1,KAMIN2} the ARPES line width, which is directly
related to Im $\Sigma$, has been analyzed for an optimally doped
Bi2212 sample ($T_{\rm c}$=90K) as a function of frequency and
temperature. In fact, it was observed that in the SC state Im
$\Sigma$ is strongly reduced below $70 ...80meV$ which compares
rather well with the drop of the scattering rate as obtained from
infrared reflectivity measurements. \cite{PUCHKOV} Moreover, with
increasing temperature there is a gradual change of Im $\Sigma$
towards a linear frequency dependence above $T_{\rm c}$. The
corresponding marginal Fermi liquid (MFL) structure of the ARPES
line shape is also discussed in Ref. \onlinecite{VARMA2}.

In order to interpolate between these two limiting self-energies
we consider the coupling of charge carriers to the following
bosonic spectrum:
\begin{equation}\label{Bw}
B(\omega)= \mbox{tanh}(\omega/(kT)) F_{\omega_0,\Gamma}(\omega)
\end{equation}
where $F_{\omega_0,\Gamma}(\omega)$ is a normalized distribution
function centered around $\omega_0$ with halfwidth $\Gamma$.
Assuming particle-hole symmetry with respect to the chemical
potential and neglecting for the moment the influence of the SC
gap \cite{NOTE}, evaluation of the $T=0$ self-energy yields
\begin{eqnarray}
\mbox{Im} \Sigma(\omega) &=& \lambda^2 N(0) \int\limits_0^{min(-\omega,\omega_c)}
                             d\nu F_{\omega_0,\Gamma}(\nu) \label{IM} \\
\mbox{Re} \Sigma(\omega) &=& -\lambda^2 N(0) \int_0^{\omega_c}
d\nu F_{\omega_0,\Gamma}(\nu) \mbox{ln}\left|\frac{\nu+\omega}{\nu-\omega}
\right| \label{RE}
\end{eqnarray}
Here $\lambda$ denotes the coupling constant, $N(0)$ corresponds to
the density of states at the chemical potential and $\omega_c$
is an upper cutoff energy for the interaction.

In the limit $\Gamma \to 0$ where $F_{\omega_0,\Gamma\to
0}(\omega)=\delta(\omega-\omega_0)$ Eqs. (\ref{IM},\ref{RE})
describe the self-energy as arising from the coupling to a single
oscillator mode. In this case one obtains the well-known
step-function behavior \cite{ENGEL} of Im $\Sigma(\omega)=1/2
\lambda^2 N(0) \Theta(|\omega|-\omega_0)$. Let us consider for
illustrative purposes at first a rectangular distribution function
$F_{\omega_0,\Gamma}(\omega)$ with halfwidth $\Gamma$ and height
$1/(2\Gamma)$ for which the self-energy can be calculated
analytically. As a result for finite but small $\Gamma<\omega_0$
the step in Im $\Sigma(\omega)$ will evolve into a linear increase
between $\omega_0-\Gamma$ and $\omega_0+\Gamma$. Finally for
$\Gamma=\omega_0$ the MFL limit is reached where Im
$\Sigma(\omega)\sim\omega$ for $0<\omega<\omega_0+\Gamma$.

\begin{figure}[tbh]
\includegraphics[width=8cm,clip=true]{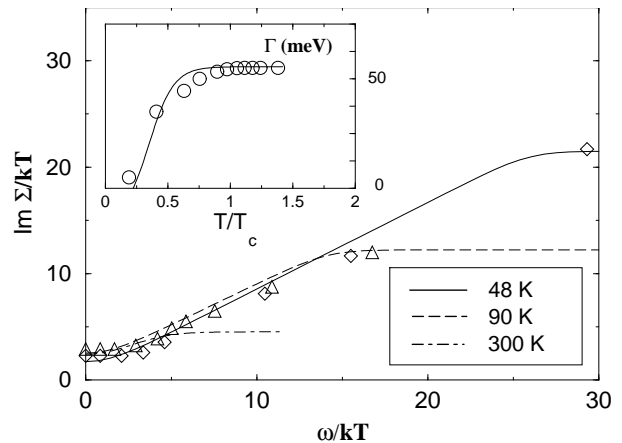}
\caption{Im $\Sigma/kT$ as a function of $\omega/kT$ for 48 K  
(solid line, diamonds), 90 K (dashed line, triangles)
and 300 K (dotted-dashed line).
Symbols are experimental data taken from Fig. 4 of Ref. \protect\onlinecite{VALLA}.
Inset: $\Gamma(T)$ as obtained from a comparison between
eq. (\ref{zfkt}) and experimental data of Ref. \protect\onlinecite{DING} (circles).}
\end{figure}

Furthermore we evaluate the quasiparticle weight
$Z=\left(1-(\partial\Sigma/\partial\omega)_{\omega=0}\right)^{-1}$
which is given by
\begin{equation}\label{zfkt}
Z=\left\lbrack1+\frac{\lambda^2 N(0)}{\Gamma}\mbox{ln}\frac{\omega_0+\Gamma}
{\omega_0-\Gamma}\right\rbrack^{-1}
\end{equation}
and vanishes logarithmically when $\Gamma$ approaches the mean
mode frequency $\omega_0$. This result can be connected with
recent ARPES experiments where the quasiparticle weight $Z$ as a
function of temperature has been extracted from the line shape at
$(\pi,0)$. \cite{DING} In fact it was found that $Z$ drops
significantly as temperature approaches $T_{\rm c}$ with a small
but nonzero $Z$ also above $T_{\rm c}$. From Eq. (\ref{zfkt}) we
thus conclude that the bosonic spectrum is governed by a
distribution function $F_{\omega_0,\Gamma}(\omega)$ with a
strongly temperature dependent halfwidth $\Gamma=\Gamma(T)$.

We have extracted $\Gamma(T)$ by comparing eq. (\ref{zfkt})
with the experimentally determined
quasiparticle weight (Fig. 4b in Ref. \onlinecite{DING}). 
The result is displayed
in the inset of Fig. 1. This plot shows that in the SC state
$\Gamma(T)$ continuously
increases and reaches the MFL limit (i.e. $\Gamma \approx
\omega_0$) for $T \approx T_c$.

With the extracted $\tanh-$like form for $\Gamma(T)$ we now evaluate
the temperature and frequency dependence of Im $\Sigma$
for parameters which also fit the real part (see below).
In Fig.1 we compare our results with ARPES data on an 
optimally doped Bi2212 sample \cite{VALLA} 
which support a scaling of Im $\Sigma$
as a function of $\omega/kT$ characteristic for a MFL. 
However, from the remarkable agreement of our fit with experiment 
as shown in Fig. 1 we conclude that the presence of a collective 
mode in the $48 K$ data cannot be excluded. We will see below that
this is also consistent with the observed temperature dependence 
of Re $\Sigma$ as required by the Kramers-Kronig transformation.
It is worth noting that $\Gamma(T)$ was obtained from a
fit to the quasiparticle weight at $(\pi,0)$ whereas the experimental data
shown in Fig. 1 are extracted from momentum distribution curves along the
nodal direction.
For large frequencies Im $\Sigma \to \lambda^2 N(0)$ and thus in our
scaled representation Fig. 1 the saturation decreases with increasing
temperature. Note that this saturation is due to the upper cutoff frequency 
of our bosonic spectrum at frequency $\omega_0+\Gamma$.

Fig. 2a displays  Re $\Sigma$ for parameters
appropriate to fit the data extracted from ARPES experiments on
an optimally doped Bi2212 sample. \cite{JOHN} 
Note that parameters are the same for both
temperatures except the broadening $\Gamma$ which in the normal
state is five times the value of that in the SC state. From our
fit we thus again conclude that the increase of temperature alone is not
sufficient in order to explain the change of  Re $\Sigma$ from
being significantly peaked at $T=70 K$ to the much broader
structure at $T=130 K$. Instead the transition from $T < T_c$ to
$T > T_c$ is accompanied by the transformation of the self-energy
from mode-like to almost MFL behavior due to the broadening of the
underlying bosonic spectrum Eq. (\ref{Bw}).

\begin{figure}[tbh]
\includegraphics[width=8.5cm,clip=true]{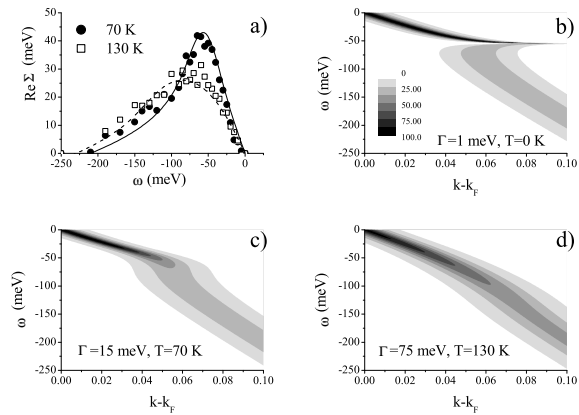}
\caption{a) Re $\Sigma(\omega)$ compared with experimental data
from Ref. \protect\onlinecite{JOHN} for 70 K (solid line, circles) and
130 K (dashed line, squares). Parameters: $\omega_0=55meV$,
$\omega_c=400meV$, $\lambda^2 N(0)=20meV$, $\Gamma(70K)=15meV$,
$\Gamma(130K)=75meV$. (b-d) Contour plot of spectral functions for
increasing values of $\Gamma$ and temperature showing the gradual
weakening of the kink feature.}
\end{figure}

In Figs. 2c-2d we show a
contour plot of the spectral function $A_k(\omega)=1/\pi
\mbox{Im}1/(\omega-\varepsilon_k-\Sigma(\omega))$ calculated for
the two values of $\Gamma$ as determined in Fig. 2a. 
The underlying bare dispersion  is linearized around the Fermi
momentum $\varepsilon_k \sim (k-k_F)$ since
experimental data for Re $\Sigma$ in Fig. 2a have been obtained by
substracting the measured nodal dispersion from a noninteracting,
linearized dispersion. \cite{JOHN} 

In addition Fig. 2b
displays the spectral function for a rather narrow boson
distribution ($\Gamma = 1meV$) at temperature T=0K. In this case
the contour plot (Fig. 2b) reveals the occurence of a break in the
dispersion separating a strongly mass enhanced low energy part
from a broad dispersive high energy feature. With increasing
$\Gamma$ (Figs. 2c, 2d) this break gradually changes into a kink
which in the MFL limit ($\Gamma \gg \omega_0$) reduces to a slight
change of slope in the dispersion. 

As already discussed in the
introduction the kink in the dispersion along the nodal direction
is now well established in a large class of cuprates.
\cite{LANZARA} Moreover ARPES experiments on underdoped Bi2212
materials (cf. Fig. 2c of Ref. \onlinecite{LANZARA}) clearly support the
existence of a 'break' also in the nodal direction corresponding
to the situation shown in Fig. 2b. Increasing doping or (and)
increasing temperature leads to a decreased change of slope (i.e.
less pronounced kink) around a characteristic binding energy (cf.
Figs. 1b,e of Ref. \onlinecite{LANZARA}). Comparing with the spectra
shown in Fig. 2b-2d all these data suggest that the charge
carriers are coupled to a bosonic spectrum which is rather sharply
peaked well below $T_{\rm c}$ but gradually broadens with
increasing temperature. In addition from the doping dependence of
the kink we conclude that the minimum $\Gamma$ is smaller in the
underdoped regime.

\section{Bilayer splitting}

Having analyzed the temperature and frequency behavior of the
bosonic spectrum we now proceed by investigating its influence on
the experimentally observed bilayer splitting in the SC and normal
state of Bi2212 materials, respectively. 

\subsection{Formalism}

Starting point is  the following hamiltonian
\begin{eqnarray}\label{H0}
H&=&\sum_{{\bf k}\sigma i} [\epsilon_{\bf k}-\mu]
c_{i,\sigma}^{\dagger}({\bf k})
c_{i,\sigma}({\bf k}) +
\sum_{{\bf k}\sigma i \ne j} t_\perp({\bf k})
c_{i,\sigma}^{\dagger}({\bf k})
c_{j,\sigma}({\bf k}) \nonumber \\
&+& \sum_{{\bf k} i} \Delta({\bf k}) \lbrack
c_{i,\uparrow}^{\dagger}({\bf k})
c_{i,\downarrow}^{\dagger}({\bf -k})+
c_{i,\downarrow}({\bf -k})
c_{i,\uparrow}({\bf k})\rbrack
\end{eqnarray}
where $,i,j=1,2$ label the planes of the bilayer system. We
restrict on an in-plane d-wave SC gap $\Delta({\bf
k})=(\Delta_0/2)[\cos(k_x)-\cos(k_y)]$ and the interlayer hopping
is parametrized as $t_\perp({\bf
k})=(t^0_{\perp}/4)[\cos(k_x)-\cos(k_y)]^2$ in agreement with LDA
\cite{LIECHT1} calculations and ARPES experiments.
\cite{FENG,CHUANG} The bare energy spectrum $\epsilon_k$ in each
plane is a tight-binding dispersion including up to fifth-neighbor
hopping. \cite{HOP} A comparison of the corresponding Fermi
surface (FS) with ARPES data from Ref. \onlinecite{FENG} is shown in the
inset of Fig.3c for an interplane matrix element
$t^0_\perp=40meV$. Diagonalization yields the four eigenvalues
$\pm E_{1,2}(k)=\pm\sqrt{(\varepsilon_k - {\mu} \pm t_\perp(k) )^2
+\Delta(k)^2}$.

The system Eq. (\ref{H0}) is now coupled to ICDW type fluctuations
via the  action
\begin{equation}\label{SEFF}
S=-\sum_q  \int_0^{\beta}d\tau_1\int_0^{\beta}d\tau_2 \sum_{i,j}
\rho^i_q(\tau_1) \chi^{ij}_q(\tau_1-\tau_2)\rho^j_{-q}(\tau_2)
\end{equation}
where the susceptibility matrix contains both intra-
($\chi_q^{\parallel}\equiv \chi_q^{i=j}$) and interlayer ($\chi_q^{\perp}
\equiv \chi_q^{i\neq j}$)
contributions. Following the approach in Ref. \onlinecite{GOETZ} we
consider an in-plane susceptibility which is factorized into an
$\omega$- and q-dependent part, i.e.
$\chi_q^{\parallel}(i\omega)=\lambda^2_{\parallel}W(i\omega)J({\bf q})$. Here,
$\lambda_{\parallel}$ denotes the intralayer coupling constant and
$W(i\omega)=-\int d\nu B(\nu) 2\nu/(\omega^2+\nu^2)$ is determined
by the bosonic spectrum Eq. \ref{Bw} for which we now consider a
lorentzian shaped distribution function
$F_{\omega_0,\Gamma}(\omega)$. $J({\bf q})$ contains the
charge-charge correlations which are enhanced at the four
equivalent critical wave vectors $(\pm q^c_x,\pm q^c_y)$.
\begin{equation}
J({\bf q})=\frac{{\cal N}}{4}
\sum\limits_{\pm q_x^c;\pm q_y^c}\frac{\gamma}{\gamma^2
+2-\cos(q_x-q_x^c)-\cos(q_y-q_y^c)}.
\end{equation}
N is a suitable normalization factor introduced to keep the total 
scattering strenght constant while varying $\gamma$.
Note that for the chosen sign convention in Eq. (\ref{SEFF}) this kind
of scattering leads to an effective attraction (in the static
limit) between holes in the same plane. However, for the
interlayer susceptibility the situation is different since the
Coulomb repulsion perpendicular to the layers is only weakly
screened and it is thus more likely to assume an effective
repulsive interaction between holes in adjacent layers. For
simplicity we therefore take in the following 
$\chi_q^{\perp}=-\chi_q^{\parallel}$ with 
$\lambda_{\parallel}=\lambda_{\perp}\equiv
\lambda$ in order to describe the physical situation where charge
enhanced regions in one layer correspond to charge depleted areas
in the other layer. In this regard the situation is analogous to
AF correlated spin fluctuations between the planes \cite{LIECHT2}
when the spin-exchange potential couples exclusively even and odd
electron states. We therefore consider our approach as complementary to
models which are based on the magnetic resonance as the relevant
collective mode. Preliminary investigations on charge-instabilities in a 
bilayer Hubbard-Holstein model \cite{GOETZ03} also give evidence for the
presence of ICDW fluctuations in the odd channel.

The leading order one-loop contributions to
the intra- $(\alpha=\parallel)$ and
interlayer $(\alpha=\perp)$ self-energies read as
\begin{equation}\label{SELF}
\underline{\underline{\Sigma}}^{\alpha}(k,i\omega)
=-\frac{\lambda^2}{\beta}
\sum_{q,ip} \chi_q^{\alpha}(ip)
\underline{\underline{\tau_z G_0^{\alpha}}}(k-q,i\omega-ip)
\underline{\underline{\tau_z}}
\end{equation}

which in turn allows for the calculation of 
$\underline{\underline{G_{}}}$ via
\begin{equation}\label{meq}
\underline{\underline{G_{}}}=\underline{\underline{G_0}}+
\underline{\underline{G_0}} \,\,\underline{\underline{\Sigma_{}}}
\underline{\underline{G_{}}}
\end{equation}
where the unperturbed Matsubara Greens function and self-energy
are $4\times4$ matrices and are given by

\begin{equation}
\underline{\underline{G_0}} \,=\, \left( \!
\begin{array}{ c c }
\underline{\underline{G_0^{\parallel}}} & \underline{\underline{G_0^{\perp}}}    \\
\underline{\underline{G_0^{\perp}}}     & \underline{\underline{G_0^{\parallel}}}  \nonumber \\
\end{array}
\right)\,\,;\,\, \underline{\underline{\Sigma}} \,=\, \left( \!
\begin{array}{ c c }
\underline{\underline{\Sigma_{}^{\parallel}}} & \underline{\underline{\Sigma_{}^{\perp}}}  \\
\underline{\underline{\Sigma_{}^{\perp}}}    & \underline{\underline{\Sigma_{}^{\parallel}}}  \nonumber \\
\end{array}
\right).
\end{equation}

Finally by inverting the $4\times4$ matrix equation  (\ref{meq})
the spectral function can be extracted from
$A_k(\omega)=Im [G_{11}(k,\omega)+G_{33}(k,\omega)]$.

\subsection{Results and Discussion}

In Fig. 3 we present results for energy distribution curves (EDC)
$I(k,\omega)=f(\omega)A(k,\omega)$ ($f(\omega)$ being the Fermi
function) in the normal (T=90K) and SC (T=10K) state and
wave-vectors are indicated in the inset of Fig. 3c. Parameters
(cf. caption to Fig. 3) were chosen in order to reproduce the
ARPES line shape features in Ref. \onlinecite{FENG}. Consider first the
normal state spectra shown in Fig. 3a. Due to the large value of
$\Gamma$ in the underlying boson spectrum the bonding band (BB)
gives rise to a broad high energy feature in the EDC at ${\bf
k}=(\pi,0)$. Together with the antibonding band (AB) quasiparticle
peak the EDC thus displays a normal state peak-dip hump structure.
The coupling to the flat boson spectrum naturally also affects the
peak-hump separation at ${\bf k}=(\pi,0)$ which therefore is
slightly larger ($\approx 90meV$) than the bare bilayer splitting
($2 t^0_\perp=80meV$). The MFL structure of both AB and BB
spectral line shape is resolved in Fig. 3c. Upon scanning from
$(\pi,0)$ towards $(\pi,\pi)$ first the AB peak goes through the
chemical potential while the BB peak shifts to lower energy and
sharpens due to the Im $\Sigma \sim \omega$ behavior. Finally
between n8 and n9 also the BB peak crosses the Fermi energy. The
AB and BB peak positions of the EDC curves from Fig. 3a are
compared in Fig. 4 with ARPES data from Ref. \onlinecite{FENG}.

\begin{figure}[bth]
\includegraphics[width=8cm,clip=true]{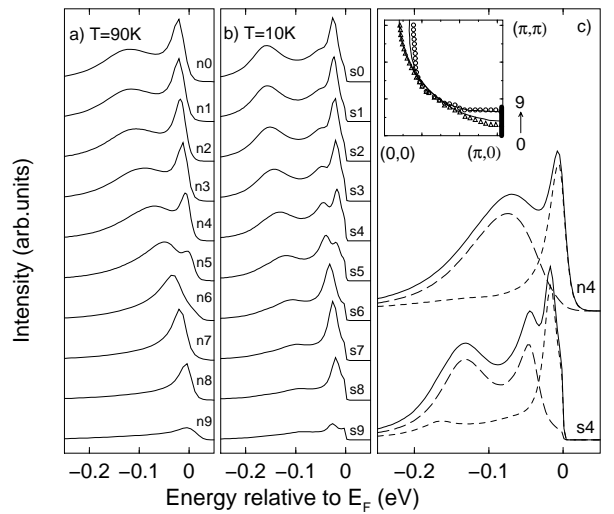}
\caption{ARPES spectra calculated for (a) normal state (T=90 K,
$\Delta_0$=0meV, $\Gamma$=100meV), (b) SC state (T=10 K,
$\Delta_0$=20meV, $\Gamma$=10meV). Spectra are taken along
$(\pi,0) \to (\pi,0.24\pi)$, and labeled from 0 to 9 as shown in  
the inset of (c). The inset displays the calculated FS in
comparison with data from Ref. \onlinecite{FENG}. (c) shows selected
spectra from (a) and (b), where dashed and dotted curves indicate
BB and AB states respectively. Parameters: $\lambda$=0.095eV,
$q_c=\pi/4$, $\omega_0$=40meV.}
\end{figure}

In the SC state the narrow bosonic distribution ($\Gamma=10meV$)
induces a strong mass enhancement for quasiparticles in the
binding energy range up to the mean mode frequency $\omega_0$.
Both AB and BB line shapes display the standard peak-dip hump
structure where the hump is more pronounced for the BB (cf. curve
s4 in Fig. 3c). The stronger renormalization of the BB band 
is due to the fact that (for our choice $\chi_q^{\perp}=-\chi_q^{\parallel}$)
the self energies of AB and BB bands 
\begin{equation}
\underline{\underline{\Sigma}}^{A,B}=-\frac{\lambda^2}{2\beta}\sum
\lbrace
(\chi_q^{\parallel}\pm\chi_q^{\perp})\underline{\underline{\tau_z
G_0^A\tau_z}}+(\chi_q^{\parallel}\mp\chi_q^{\perp})
\underline{\underline{\tau_z G_0^B\tau_z}}\rbrace
\end{equation}
are determined by fermionic excitations of
BB and AB bands, respectively. Since the van-Hove singularity of the
BB band is well separated from the chemical potential 
this results in a small AB self-energy and thus a relatively weak
AB hump feature. In contrast the closeness of the AB van-Hove
singularity to $\mu$ induces large BB self-energy effects and
therefore a more pronounced BB hump structure.
 
\begin{figure}[tbh]
\includegraphics[width=8cm,clip=true]{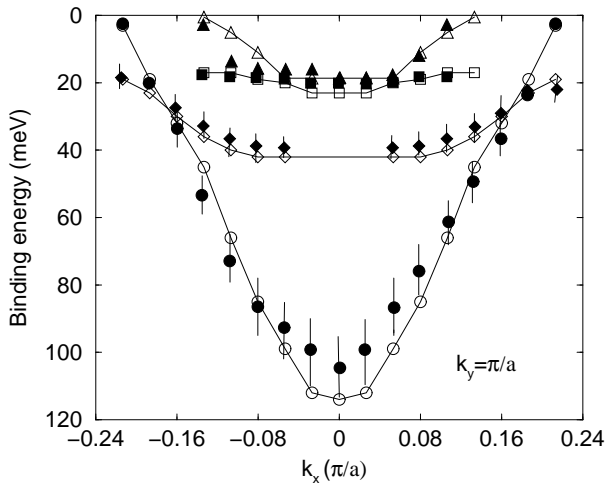}
\caption{Dispersion extracted from Fig.3 (open symbols); Filled
symbols indicate experimental data from Ref. \onlinecite{FENG}. Circles
and triangles correspond to BB and AB normal state peaks,
respectively; Diamonds and squares refer to BB and AB
quasiparticle peaks in the SC state.}
\end{figure}

The SC low energy 'quasiparticle' peaks are confined
to energies smaller than $\omega_0$ and the corresponding
quasiparticle weight decreases with increasing binding energy. As
a consequence the quasiparticle peak of the bonding band is
strongly suppressed at ${\bf k}=(\pi,0)$ (curve s0 in Fig. 3b) and
reemerges when the bare energy of the bonding band approaches
$\omega_0$ (curve s2 in Fig. 3b). In Fig. 4 where the low energy
peak positions are shown as a function of $k_x$ the dispersion of
the BB therefore seems to be interrupted around $(\pi,0)$ in
agreement with the experimental data of Ref. \onlinecite{FENG}.
Moreover, Fig. 4 displays one of the main results of this paper.
Whereas in the normal state the bilayer splitting is around
$90meV$  the low energy peaks in the SC state are separated by
$\approx 20meV$ only. As already discussed above the decreased
bilayer splitting in the SC state is only due to the sharpening of
the collective mode at low temperatures and not influenced by any
other parameter in our calculation.

\section{Conclusion}

To conclude, we have shown that the self-energy as extracted from
ARPES experiments can be consistently described within a model
where the charge carriers are coupled to a bosonic distribution
centered around some mean frequency $\omega_0$. The width $\Gamma$
of this spectrum turns out to be strongly temperature dependent as
can be concluded from the temperature dependent quasiparticle
weight (cf. inset to Fig. 1), the evolution of the kink along the node
(Figs. 2b-2d) and the corresponding change of Re $\Sigma$ in Fig.
2a with temperature. 
Within the ICDW scenario the associated
scattering can account for the experimentally observed
\cite{DESSAU} appearance of an additional peak-dip hump structure
below T$_{\rm c}$ near the $(\pi,0)$ points of the BZ (curve s4 in
Fig. 3c). Moreover, the strong QP mass enhancement below T$_{\rm
c}$ due to the small value of $\Gamma$ leads to an apparent
reduction of the bilayer splitting and a suppression of BB
spectral weight around the $(\pi,0)$ points (Fig. 4). 

Note that phonons cannot account for the temperature
dependence of the collective mode 
although their linewidth may undergoe some sharpening upon
entering the SC state. For temperatures well below the phonon frequency 
one can estimate  the linewidth \cite{ALLEN}
from $\Gamma\approx \frac{\pi}{2}N(0)\lambda \omega_0^2$
which yields $\Gamma << \omega_0$ also for large coupling
$\lambda\approx 1 ... 2$. This is not enough to account for the strong
damping as discussed above. 

On the other hand the temperature dependent width of the bosonic spectrum 
can at least be made plausible within the ICDW scenario. 
At low temperatures the large intensity CDW mode at 
energy $\omega_0$ and wave vecors $(\pm q_c,\pm q_c)$ is part of
a band of excitations which for $q\to 0$ corresponds to the 
phason mode of the system. It is known \cite{GRUENER} that the effective
mass of the phason strongly increases with temperature
due to the decreasing CDW gap inducing a significant 
broadening of the bandwidth of charge excitations.
In fact, preliminary results obtained for the stripe phase
of the Hubbard model \cite{VARL} near the CDW transition 
have already revealed a rather
continuous frequency spectrum in the charge susceptibility
centered around $q_c$ corresponding to the stripe periodicity.
However, a more detailed investigation of this temperature dependence
remains an interesting issue for future work.

\section{Acknowledgments}

We would like to thank M. Grilli for helpful comments and a critical
reading of the manuscript.


\begin{thebibliography}{9}
\bibitem{NOTE} Most of the ARPES data analyzed in the first part
               are for k-states along the diagonals where the SC gap
               vanishes anyhow.
\bibitem{HOP} In order to reproduce the FS we choose $t_1=-590,
              t_2=181, t_3=-48.5, t_4=-19.5, t_5=11$ (in meV).
\end{thebibliography}
\end{document}